\documentclass[a4paper,UKenglish,cleveref, autoref, thm-restate]{oasics-v2021}

\usepackage{multicol}

\title{Refactoring = Substitution + Rewriting} 
\subtitle{Towards generic, language-independent refactorings}

\author{Simon Thompson\footnote{Corresponding author}}{School of Computing, University of Kent, Canterbury, UK \and Faculty of Informatics, Eötvös Loránd University, Budapest, Hungary}{s.j.thompson@kent.ac.uk}{https://orcid.org/0000-0002-2350-301X}{With support from the UK Engineering and Physical Sciences Research Council awards EP/N028759/1, EP/C524969/1, GR/R75052/01 and EP/T014512/1.}

\author{Dániel Horpácsi}{Faculty of Informatics, Eötvös Loránd University, Budapest, Hungary}{daniel-h@elte.hu}{https://orcid.org/0000-0003-0261-0091}{
With support from the NRDI Fund of Hungary and financed under the Thematic Excellence Programme TKP2020-NKA-06 (National Challenges Subprogramme) funding scheme.
}

\authorrunning{S. Thompson and D. Horpácsi}

\Copyright{Simon Thompson and Dániel Horpácsi}



\category{} 

\relatedversion{} 

\nolinenumbers 

\EventEditors{John Q. Open and Joan R. Access}
\EventNoEds{2}
\EventLongTitle{42nd Conference on Very Important Topics (CVIT 2016)}
\EventShortTitle{CVIT 2016}
\EventAcronym{CVIT}
\EventYear{2016}
\EventDate{December 24--27, 2016}
\EventLocation{Little Whinging, United Kingdom}
\EventLogo{}
\SeriesVolume{42}
\ArticleNo{23}

\begin{document}

\maketitle

\begin{abstract}
Eelco Visser's work has always encouraged stepping back from the particular to look at the underlying, conceptual problems. 

In that spirit we present an approach to describing refactorings that abstracts away from particular refactorings to classes of similar transformations, and presents an implementation of these that works by substitution and subsequent rewriting. 

Substitution is language-independent under this approach, while the rewrites embody language-specific aspects. Intriguingly, it also goes back to work on API migration by Huiqing Li and the first author, and sets refactoring in that general context. 
\end{abstract}

\section{Introduction}
\label{sec:intro}

Our subject here is not new: Eelco Visser initiated a discussion of language-independence transformations~\cite{EVLangIndep} in the millennium year; it is a pleasure and an honour to join the conversation that he began.

Refactoring tools are a particularly sensitive part of a programmer's toolkit, since they can make large-scale modifications to code, and yet are expected not to change the observable behaviour of the system. Users therefore need to be given assurance about the safety of using a tool. Assurance generally comes in two complementary forms, through verification and through architecture.

Arguments can be made about the \emph{correctness} of the transformations made. These can be black box, checking the original against the refactored code without examining how the transformation is performed. At minimum this is achieved by regression testing, but can be augmented by checking equivalence more generally~\cite{PEQCHECK,PEQtest}. Looking inside the implementation it is also possible -- at least in principle -- to prove the correctness of that transformation~\cite{nik-refac}.

These checks will apply to particular \emph{instances} of a refactoring in the case of regression testing, while verification should establish the correctness of the transformation in itself, that is for \emph{all} possible instances. Testing can also be used in the latter case, e.g. by generating random transformations of an arbitrary system, and testing the `old' code against the `new' with random inputs~\cite{RandomtestEW09}.

An alternative source of assurance is in the \emph{architecture} of the refactoring tool itself. At its heart, any refactoring tool works by transforming a complex data structure (AST or database) representing a program into a related structure. With no further thought, each refactoring can be constructed anew, by means of an \emph{ad hoc} recursive function. But this can be improved. Firstly, following Eelco Visser's work~\cite{EVLangIndep}, it is possible to take a  higher-level view of the way the data structure is traversed, using a \emph{strategic} approach in something like Stratego~\cite{stratego}, Strafunski~\cite{Strafunski} and related systems.

Secondly, we can see a commonality between the refactorings themselves, and this is the approach we outline here. With this perspective, a class of refactorings can all be performed with a single implementation. This, in turn, reduces the burden on users wishing to assure themselves of the soundness of the implementation, and indeed has positive consequences for formal verification of the refactoring transformations too.

In the remainder, we first introduce the separation of generic and specific elements of refactoring definitions in \S\ref{sec:refac-sub-rewrite}. Then in \S\ref{sec:generic} we argue that the generic parts in many cases can be made language-independent. Finally, \S\ref{sec:related} discusses some related work and \S\ref{sec:conclusion} concludes.

\section{Refactoring = Substitution + Rewriting}
\label{sec:refac-sub-rewrite}

A class of refactorings in Erlang are concerned with the definition and subsequent use of functions, including, among others, renaming and generalisation. In this section we show how these can be described in a common way, and the implications of this for verification.

\subsection{Function renaming}

Consider the example of function renaming. 

\begin{multicols}{2}

\noindent
Before renaming
\begin{verbatim}
f(X) -> X+1.
g(Y) -> f(Y+2) - f(Y-2).
\end{verbatim}
\columnbreak

\noindent
After renaming
\begin{verbatim}
h(X) -> X+1.
g(Y) -> h(Y+2) - h(Y-2).
\end{verbatim}
\end{multicols}
\noindent
The transformation is described by showing how the function \emph{definition} is changed, and also explaining how to change each \emph{use} of the function. We do that thus:
\begin{multicols}{2}

\noindent
Define the modified function:\\  
\verb^h(X) -> X+1.^

\columnbreak
\noindent
Implement the old function using the new:\\  \verb+f = fun(X) -> h(X) end+
\end{multicols}
\noindent
How does this describe the refactoring? We can use the implementation of the old function in terms of the new to give us the new code, in a series of steps, thus:
\begin{verbatim}
g(Y) -> f(Y+2) - f(Y-2).
-- by substitution giving
g(Y) -> (fun(X) -> h(X) end)(Y+2) - (fun(X) -> h(X) end)(Y-2).
-- and by rewriting (beta-reduction, here)
g(Y) -> h(Y+2) - h(Y-2).
\end{verbatim}
Rewriting stops at this point: we don't want, or need, to inline \texttt{h}, since we want to be faithful to the original program that contains a function call to \texttt{f} here.

It is important to note that this approach works for other uses of the function \verb+f+, including as \verb+fun+ arguments to higher-order functions, and, with some preliminary eta-expansion \footnote{ Transforming \texttt{fun f/1} into \texttt{fun(X)->f(X) end}.}, in calls to \verb+spawn+ the function.

\subsection{Function generalisation}
Now we look at a second example, function generalisation, and see that it fits the same pattern.

\begin{multicols}{2}
\noindent
Before generalisation
\begin{verbatim}
f(X) -> X+3.
g(Xs) -> lists:map(fun f/1,Xs).

\end{verbatim}
\columnbreak

\noindent
After generalisation
\begin{verbatim}
f(X,Y) -> X+Y.
g(Xs) -> 
    lists:map(fun(X)-> f(X,3) end, Ys).
\end{verbatim}
\end{multicols}

\noindent
This transformation is described by showing how the function \emph{definition} is changed, and also explaining how to change each \emph{use} of the function. We do that thus:
\begin{multicols}{2}

\noindent
Define the modified function:\\  
\verb^f(X,Y) -> X+Y.^

\columnbreak
\noindent
Implement the old function using the new:\\  \verb+f = fun(X)-> f(X,3) end+
\end{multicols}
\noindent
How does this describe the refactoring? We can use the implementation to give us the new code, in a series of steps, thus:
\begin{verbatim}
g(Xs) -> lists:map(fun f/1,Xs).
-- by substitution giving
g(Xs) -> lists:map(fun(X)-> f(X,3) end, Ys).
-- after which no rewriting is necessary
\end{verbatim}
The implementation of the old function in terms of the new is denoted by \texttt{=} rather than \texttt{->} to emphasise that this is a semantic equivalence rather than program code defining a function, since the LHS refers to the `old' version of the code and the RHS to the `new'.\footnote{It is a peculiarity of Erlang that the two versions of \texttt{f} can co-exist, as functions with the same name but different arity are considered to be distinct.}

As earlier, this approach will handle `regular' function applications, in which the \texttt{fun} expression will be removed by rewriting, as well as calls to \verb+apply+ and \verb+spawn+.

\subsection{Other function-oriented examples}

Other examples include function unfolding, reordering and regrouping of arguments, adding or removing an argument. We leave it to the reader to verify this. In each case, the general pattern is to present the \textbf{new definition} and to describe how to \textbf{implement the old function using the new.}

We examine other kinds of refactoring in Section~\ref{other-kinds}.

\subsection{Verification}

What do we need to establish for the transformed code to be equivalent to the original? The verification factors into two parts:

\begin{itemize}
\item
For \textbf{each specific refactoring} it is necessary to ensure that the `old' function is implemented correctly in terms of the `new'. Specifically, the replacement becomes a \emph{proof obligation}. In the case of renaming, we require that \texttt{f} has the same behaviour as
\begin{verbatim}
    fun(X) -> h(X) end
\end{verbatim}
when \texttt{h} is defined thus:
\begin{verbatim}
    h(X) -> X+1.
\end{verbatim}
Similarly for other refactorings.
\item
On the other hand, \textbf{every refactoring} also depends on the correctness of the rewriting rules, such as beta-reduction, eta-expansion, removal of syntactic sugar, etc., which are applied to `tidy up' the resulting code in each case.
\end{itemize}

\section{Towards generic, language-independent refactorings}
\label{sec:generic}

Refactoring has a very different character in different programming languages; to take one example, \cite{SCAM06} compares refactoring in two functional programming languages: Haskell and Erlang. Because of this, the first author was always sceptical about a \emph{language-independent} approach to refactoring. In this Section we argue that our approach of substitution and rewrite allows us to split refactorings into language-independent and language-dependent parts.

At the \textbf{language independent} level is a concept like \emph{function application}; function applications can be transformed by \emph{defining the old function in terms of the new}, as described earlier. On the other hand, the \textbf{particular} form of function application in different languages varies widely, for example.

\begin{itemize}
    \item In \textbf{Haskell} function applications can be infix \verb+x `f` y+ as well as prefix \verb+f x y+, and functions -- prefix or infix-- can also be partially applied, as in the expressions \verb+map (f x) xs+ and \verb+map (x `f`) xs+.
    \item While \textbf{Erlang} does not contain infix  function or partial applications, functions are passed as arguments using the `function/arity' idiom \verb+ fun/N+, but also be can be referenced by \emph{atoms} in some special functions, such as \texttt{spawn}.
\end{itemize}

\noindent 
These differences can be dealt with by means of rewriting, as we saw with Erlang earlier. Consider the Haskell example 
\begin{verbatim}
f x y = x+y

g z xs = map (f z) xs
\end{verbatim}
where the order of arguments to \verb+f+ is reversed, so that the original \verb+f+ is implemented in terms of the new thus \verb+\x y -> f y x+.
Applying this transformation to the definition of \verb+g+ gives
\begin{verbatim}
g z xs = map (f z) xs
-- substituting the definition of f
g z xs = map ((\x y -> f y x) z) xs
-- by beta reduction
g z xs = map (\y -> f y z) xs
\end{verbatim}
This leaves a lambda (`\verb+\+') in the refactored expression, but this is unavoidable. While this example might have been handled better using the \texttt{flip} function, this approach generalises to any permutation (or tupling) of the arguments by introducing the appropriate, unnamed, equivalent of \texttt{flip} as the lambda expression.

If, on the other hand, we had renamed  \verb+f+ to  \verb+h+, the redefinition of  \verb+f+ would be \verb+\x y -> h x y+, and the refactoring would completely eliminate the introduced lambda thus:

\begin{verbatim}
g z xs = map (f z) xs
-- substituting the definition of f
g z xs = map ((\x y -> h x y) z) xs
-- by beta reduction
g z xs = map (\y -> h z y) xs
-- by eta reduction
g z xs = map (h z) xs
\end{verbatim}

It is also possible to accommodate infix operations into this framework too. One option is to recognise \verb+`f`+ as a function syntax; alternatively, and preferably, we can \emph{pre-process} the code prior to substitution. In this case we proceed thus:

\begin{verbatim}
g z xs = map (z `f`) xs
-- replacing the infix "syntactic sugar"
g z xs = map (infix f z) xs
-- substituting the definition of f
g z xs = map (infix (\x y -> h x y) z) xs
-- by eta reduction
g z xs = map (infix (\x -> h x) z) xs
-- by eta reduction
g z xs = map (infix h z) xs
-- reintroducing the infix "syntactic sugar"
g z xs = map (z `h`) xs
\end{verbatim}

Language dependence can extend beyond syntactic sugar. For example, in Ocaml function names can appear in signatures and as arguments to functors, where parameters are identified by name rather than position. This impacts the way in which the scope of a renaming refactoring is identified, as explained in~\cite{Rotor-PLDI}.

\section{Discussion}
\label{sec:discussion}

The approach discussed here is based on some assumptions, and so has some advantages, as well as some limitations. We discuss these in more detail now.

\subsection{Local \emph{vs} global} 

Many refactorings are \emph{local}, in the sense of being applied at a single point, such as a replacement of a double list traversal \verb+map f. map g+ by a single one \verb+map (f.g)+, but it is \emph{global} refactorings, whose effect might span multiple sites within multiple modules, that are more problematic to implement and to review, e.g.\ in a pull request. 

We have concentrated on global refactorings here for that reason, but a rewriting approach plainly works well for implementing local refactorings too, as shown by the retrie~\cite{retrie} tool for Haskell. On the other hand, it is difficult to see most local refactorings as anything other than language specific.

\subsection{Recursion \emph{vs} iteration}

How might the replacement of recursion by iteration be seen in this framework? To encapsulate recursion in general would require some kind of template language, but then an arbitrary recursion cannot be replaced by iteration. Once the recursion has a stylised form, this can be encapsulated in a combinator, as in 

\begin{verbatim}
diffs xs = foldr (-) 0 xs
\end{verbatim}

\noindent
and then a transition to an iterative form can be given thus

\begin{verbatim}
diffs xs = foldl (flip (-)) 0 (reverse xs)
\end{verbatim}

\noindent
Further transformation can render the list reverse in an iterative way too. This has been expressed in the syntax of Haskell, but all functional languages contain cognates of lists and folding operations, and so, arguably, it has a language-independent aspect.

\subsection{Other kinds of refactoring}
\label{other-kinds}

A similar approach can be taken to \emph{constructor-based} refactorings in languages like Haskell and OCaml. A constructor is like a function, except that it can be used on the `left hand side' of definitions in pattern matches, and this requires some limited form of rewriting on patterns to implement.

There are limits to the approach described here. For example, `folding' function definitions, i.e.\ replacing instances of a function body by a call to the function necessitate replacing (an instance of) a complex expression, rather than a single term. In the short term, we aim more clearly to articulate the scope and limits of the approach.

\section{Related work}
\label{sec:related}

\subsection{Language-independence and genericity}
\label{subsec:lang-indep}

The questions of language-independence and genericity for refactorings have been addressed before. Indeed, Eelco Visser initiated a discussion of this in the millennium year in \emph{Language Independent Traversals for Program Transformation}~\cite{EVLangIndep}, which described how strategic, traversal-based programming could achieve transformations such as change in bound variable names across functional and OO languages that could be subsets of Haskell and C++ respectively. 
Shortly after this L\"{a}mmel's \emph{Towards Generic Refactoring}~\cite{towardsGeneric} took this explicitly to the example of function/abstraction extraction with an approach that performs a conceptual analysis of the categories of transformations and pre-conditions that are necessary for a generic treatment of a refactoring. 

This approach is flexible and comprehensive, but it lacks completeness. While it can encompass the generic features that occur in multiple languages, and indeed adapt e.g.\ to the transition between expression- and statement-oriented languages, it does not support the particularities of different languages, such as operator sections in Haskell or the use of atoms to denote functions in Erlang, that our approach can handle.

While we have argued that our approach supports a degree of language independence, we would not claim that it directly supports multi-language refactoring~\cite{hurdles}, since that requires not only awareness of the separate semantics of a number of languages, but also their semantic interactions.

\subsection{Verification of refactorings}
\label{subsec:verification}

A powerful approach to ensuring the correctness of refactorings is to ensure that they meet the set of constraints that embody (aspects of) the semantics of the programming language being refactored. This insight was first presented by Tip and colleagues in the context of preserving \emph{type constraints}~\cite{TipTypeConstraints}, and elaborated for Java by de Moor and Schaefer~\cite{SchaeferDeMoor}. Steimann~\cite{constraint-based-refactoring} presents a general theory of constraint-based refactoring, and outlines a program in which correct-by-construction tools can be built on top of constraint-based presentations of programming language semantics.

Pioneering work by Kniesel and Koch~\cite{StaticComposition} examines the way that correct refactorings can be built by composing simpler parts that themselves preserve behaviour, or can be verified separately. Our approach is related, but differs in that different instances of the same refactoring will involve different rewrites, depending on the context of the instance: the composition is thus, in a sense, dynamic. 

\subsection{API migration}
\label{subsec:api}

When an API is upgraded it can be taken for granted that the new API should afford all the functionality provided by the old version; this can be made concrete in an \emph{adapter module} that defines the old in terms of the new. 

While adaptation is enough to ensure that the client system continues to work, it has disadvantages. If an API evolves continually, then a series of adapter modules will stack up, and even in the case of a single adaptation, the code will be neither idiomatic nor natural. One approach to this is to generate transformations from the replacement code~\cite{AutoGenAPIrefac}, which ensures that the explicit wrapper code disappears. This mechanism is extended by our approach, outlined in~\cite{ASE12}, where the replacement code is subsequently simplified by rewriting, e.g.\ removing case expressions when they can be resolved, or exception-handling code when that is unnecessary.

This adaptation can be complex, however, particularly in the case of object-oriented programming, and especially when the migration is from one API to another, unrelated one.  L\"{a}mmel and colleagues outline this in a case study~\cite{SwingToSWT} of evolving a system, while providing a broad overview of previous approaches, as well as in this general exploration of two XML case studies here~\cite{bartolomei2010study}.

It turns out that the approach we outline in this paper can be seen as a particular case of the work presented in~\cite{ASE12}, viewing each refactoring as an evolution of an API for those aspects of the code that has changed.

\subsection{Refactoring schemes}
\label{subsec:schemes}

The approach explained in this paper can be seen as a variant of the \emph{refactoring schemes} proposed in~\cite{Horpacsi_2017}. In particular, the examples given in Section~\ref{sec:refac-sub-rewrite} instantiate the function refactoring scheme, which can be understood as a strategy that changes function entities in a program by applying rewrite rules to the definition and to the references of the function. When restricting the rewrite rules to only rewrite the name of the function in the reference, the rewrite step does not perform pattern matching and thus it becomes a simple substitution.

\section{Conclusions and future work}
\label{sec:conclusion}

We have advanced an argument that it is possible to view general refactorings as having a language-independent component, described in the language of function application, and a language-dependent component, materialised by a set of language-specific rewrite rules. This description re-frames earlier work of ours on refactoring for API evolution and language schemes.

We have experimental implementations of the general function refactoring introduced in Section \ref{sec:refac-sub-rewrite} in the Wrangler~\cite{LiThompsonErlang08} refactoring tool for Erlang, where it is materialised as an Erlang \texttt{behaviour}, and in the Rotor~\cite{Rotor} refactoring tool for OCaml. It is a short term goal to finalise and deploy these implementations, as well as articulating the scope and limits of the approach itself.

This work forms part of a longer-term project to build high assurance refactorings. Earlier work on this has concentrated on a formal treatment of (re-)naming in OCaml~\cite{Rotor-PLDI}, and a formalisation of the semantics of Erlang~\cite{formal-seq-erlang}.

We are very grateful to the referees for their feedback, and in particular their encouragement to contextualise the work more thoroughly.

\bibliography{EVCS-Thompson-Horpacsi}

\begin{thebibliography}{10}

\bibitem{SwingToSWT}
Thiago~Tonelli Bartolomei, Krzysztof Czarnecki, and Ralf L{\"{a}}mmel.
\newblock Swing to {SWT} and back: Patterns for {API} migration by wrapping.
\newblock In Radu Marinescu, Michele Lanza, and Andrian Marcus, editors, {\em
  26th {IEEE} International Conference on Software Maintenance {(ICSM} 2010),
  September 12-18, 2010, Timisoara, Romania}, pages 1--10. {IEEE} Computer
  Society, 2010.
\newblock \href {https://doi.org/10.1109/ICSM.2010.5610429}
  {\path{doi:10.1109/ICSM.2010.5610429}}.

\bibitem{bartolomei2010study}
Thiago~Tonelli Bartolomei, Krzysztof Czarnecki, Ralf L{\"a}mmel, and Tijs Van
  Der~Storm.
\newblock {Study of an API migration for two XML APIs}.
\newblock In {\em International Conference on Software Language Engineering},
  pages 42--61. Springer, 2010.

\bibitem{formal-seq-erlang}
P{\'e}ter Bereczky, D{\'a}niel Horp{\'a}csi, and Simon Thompson.
\newblock {A Proof Assistant Based Formalisation of a Subset of Sequential Core
  Erlang}.
\newblock In Aleksander Byrski and John Hughes, editors, {\em {Trends in
  Functional Programming}}, pages 139--158, Cham, 2020. Springer International
  Publishing.

\bibitem{stratego}
Martin Bravenboer, Karl~Trygve Kalleberg, Rob Vermaas, and Eelco Visser.
\newblock {Stratego/XT 0.17. A language and toolset for program
  transformation}.
\newblock {\em {Sci. Comput. Program.}}, 72, 2008.

\bibitem{RandomtestEW09}
D{\'a}niel Drienyovszky, D{\'a}niel Horp{\'a}csi, and Simon Thompson.
\newblock {QuickChecking Refactoring Tools}.
\newblock In Scott~Lystig Fritchie and Konstantinos Sagonas, editors, {\em
  {Erlang'10: Proceedings of the 2010 ACM SIGPLAN Erlang Workshop}}, pages
  75--80. ACM SIGPLAN, 2010.

\bibitem{Horpacsi_2017}
D{\'{a}}niel Horp{\'{a}}csi, Judit K{\H{o}}szegi, and Zolt{\'{a}}n
  Horv{\'{a}}th.
\newblock {Trustworthy Refactoring via Decomposition and Schemes: A Complex
  Case Study}.
\newblock {\em Electronic Proceedings in Theoretical Computer Science},
  253:92--108, Aug 2017.
\newblock \href {https://doi.org/10.4204/eptcs.253.8}
  {\path{doi:10.4204/eptcs.253.8}}.

\bibitem{PEQCHECK}
Marie-Christine Jakobs.
\newblock {PEQCHECK: Localized and Context-aware Checking of Functional
  Equivalence}.
\newblock In {\em {2021 IEEE/ACM 9th International Conference on Formal Methods
  in Software Engineering (FormaliSE)}}, pages 130--140, 2021.
\newblock \href {https://doi.org/10.1109/FormaliSE52586.2021.00019}
  {\path{doi:10.1109/FormaliSE52586.2021.00019}}.

\bibitem{PEQtest}
Marie-Christine Jakobs and Maik Wiesner.
\newblock {PEQtest: Testing Functional Equivalence}.
\newblock In Einar~Broch Johnsen and Manuel Wimmer, editors, {\em {Fundamental
  Approaches to Software Engineering}}, pages 184--204, Cham, 2022. Springer
  International Publishing.

\bibitem{StaticComposition}
Günter Kniesel and Helge Koch.
\newblock Static composition of refactorings.
\newblock {\em Science of Computer Programming}, 52(1):9--51, 2004.
\newblock Special Issue on Program Transformation.
\newblock \href {https://doi.org/doi:10.1016/j.scico.2004.03.002}
  {\path{doi:doi:10.1016/j.scico.2004.03.002}}.

\bibitem{towardsGeneric}
Ralf L\"{a}mmel.
\newblock {Towards Generic Refactoring}.
\newblock In {\em Proceedings of the 2002 ACM SIGPLAN Workshop on Rule-Based
  Programming}, RULE '02, page 15–28, New York, NY, USA, 2002. Association
  for Computing Machinery.
\newblock \href {https://doi.org/10.1145/570186.570188}
  {\path{doi:10.1145/570186.570188}}.

\bibitem{SCAM06}
Huiqing Li and Simon Thompson.
\newblock {A Comparative Study of Refactoring Haskell and Erlang Programs}.
\newblock In M.~Di Penta and L.~Moonen, editors, {\em {Source Code Analysis and
  Manipulation, SCAM'06}}, 2006.

\bibitem{ASE12}
Huiqing Li and Simon Thompson.
\newblock {Automated API Migration in a User-Extensible Refactoring Tool for
  Erlang Programs}.
\newblock In Tim Menzies and Motoshi Saeki, editors, {\em {Automated Software
  Engineering, ASE'12}}. IEEE Computer Society, 2012.

\bibitem{LiThompsonErlang08}
Huiqing Li, Simon Thompson, Gy\"{o}rgy Orosz, and Melinda T\"{o}th.
\newblock {R}efactoring with {W}rangler, updated.
\newblock In {\em {ACM SIGPLAN Erlang Workshop 2008, Victoria, British
  Columbia, Canada}}, 2008.

\bibitem{Strafunski}
Ralf Lämmel and Joost Visser.
\newblock {A Strafunski Application Letter}.
\newblock {\em Information and Computation/information and Control - IANDC},
  pages 357--375, 01 2003.
\newblock \href {https://doi.org/10.1007/3-540-36388-2_24}
  {\path{doi:10.1007/3-540-36388-2_24}}.

\bibitem{AutoGenAPIrefac}
Jeff~H. Perkins.
\newblock {Automatically Generating Refactorings to Support API Evolution}.
\newblock {\em SIGSOFT Softw. Eng. Notes}, 31(1):111–114, Sep 2005.
\newblock \href {https://doi.org/10.1145/1108768.1108818}
  {\path{doi:10.1145/1108768.1108818}}.

\bibitem{retrie}
{Retrie, a powerful, easy-to-use codemodding tool for Haskell.}, 2020.
\newblock \url{https://github.com/facebookincubator/retrie}.

\bibitem{Rotor-PLDI}
Reuben Rowe, Hugo F{\'e}r{\'e}e, Simon Thompson, and Scott Owens.
\newblock {Characterising renaming within Ocaml’s module system: theory and
  implementation}.
\newblock In {\em {Proceedings of the 40th ACM SIGPLAN Conference on
  Programming Language Design and Implementation}}, pages 950--965, 2019.

\bibitem{Rotor}
Reuben Rowe, Hugo Férée, Simon Thompson, and Scott Owens.
\newblock {ROTOR: A Tool for Renaming Values in OCaml's Module System}.
\newblock In {\em {2019 IEEE/ACM 3rd International Workshop on Refactoring
  (IWoR)}}, pages 27--30, 2019.
\newblock \href {https://doi.org/10.1109/IWoR.2019.00013}
  {\path{doi:10.1109/IWoR.2019.00013}}.

\bibitem{SchaeferDeMoor}
Max Schaefer and Oege de~Moor.
\newblock {Specifying and Implementing Refactorings}.
\newblock {\em SIGPLAN Not.}, 45(10):286–301, oct 2010.
\newblock \href {https://doi.org/10.1145/1932682.1869485}
  {\path{doi:10.1145/1932682.1869485}}.

\bibitem{hurdles}
Hagen Schink, Martin Kuhlemann, Gunter Saake, and Ralf Lämmel.
\newblock {Hurdles in Multi-language Refactoring of Hibernate Applications.}
\newblock In {\em ICSOFT 2011 - Proceedings of the 6th International Conference
  on Software and Database Technologies}, volume~2, pages 129--134, 01 2011.

\bibitem{constraint-based-refactoring}
Friedrich Steimann.
\newblock {Constraint-Based Refactoring}.
\newblock {\em ACM Trans. Program. Lang. Syst.}, 40(1), Jan 2018.
\newblock \href {https://doi.org/10.1145/3156016} {\path{doi:10.1145/3156016}}.

\bibitem{nik-refac}
Nik Sultana and Simon Thompson.
\newblock {M}echanical {V}erification of {R}efactorings.
\newblock In {\em {Workshop on Partial Evaluation and Program Manipulation}}.
  ACM SIGPLAN, 2008.

\bibitem{TipTypeConstraints}
Frank Tip, Robert~M. Fuhrer, Adam Kie\.{z}un, Michael~D. Ernst, Ittai Balaban,
  and Bjorn De~Sutter.
\newblock {Refactoring Using Type Constraints}.
\newblock {\em ACM Trans. Program. Lang. Syst.}, 33(3), may 2011.
\newblock \href {https://doi.org/10.1145/1961204.1961205}
  {\path{doi:10.1145/1961204.1961205}}.

\bibitem{EVLangIndep}
Eelco Visser.
\newblock {Language Independent Traversals for Program Transformation}.
\newblock In J.~Jeuring, editor, {\em Proceedings of the Workshop on Generic
  Programming (WGP2000), Ponte de Lima, Portugal}, July 2000.
\newblock Technical Report, Department of Information and Computing Sciences,
  Universiteit Utrecht.

\end{thebibliography}

\end{document}